# Combined effect of Fluid Rheology and Surface Modification on Eletrokinetic Energy Generation through Finite Length Microchannel

Aditya Patwari[1,2], Avinash Kumar[2], Chirodeep Bakli[2] and Suman Chakraborty[1,3]

[1]Advanced Technology Development Centre, Indian Institute of Technology Kharagpur, Kharagpur-721302, India
[2]Thermofluidics and Nanotechnology for Sustainable Energy Systems Laboratory, School of Energy Science and Engineering, Indian Institute of Technology Kharagpur, Kharagpur-721302, India
[3]Department of Mechanical Engineering, Indian Institute of Technology Kharagpur, Kharagpur-721302, India

**ABSTRACT**

Electrokinetic energy conversion provides a scheme for energy harvesting and storage for on-chip applications. However, the major drawback of electrokinetic energy conversion is its low conversion efficiency. Researchers are in a quest to find ways to improve this efficiency. With the same motive, we investigated the generation of streaming potential by applying surface modification and employing a non-Newtonian fluid to flow through the microchannel under constant pressure difference across its ends. Shear-thickening liquids tend to lessen electrokinetic effects, whereas shear-thinning liquids favour them. Also, having superhydrophobic surfaces improve the magnitude of generated streaming potential. We examine the combined effect of fluid rheology and surface modification on electrokinetic energy generation. We have learned intriguing insights about using non-Newtonian fluid in hydrophobic microchannels as an outcome of our combined research. Hydrophobic surfaces do not enhance the efficiency for a fluid with below a power law index of 0.7. The findings of this research can be used towards the selection of fluid-substrate combination that will optimize electrokinetic power generation efficiency.

## 1. INTRODUCTION

Micro/nanofluidics research is getting more and more popular for its applications to everyday objects like point-of-care devices [1], inkjet printers [2], electronic cooling [3], [4] and lab-on-chip [5] devices that can shrink a whole laboratory down to a few square inches. From the point of view of small scale power harvesting and sensing technologies, micro- and nanoscale "Electrokinetics" has recently got a lot of research attention. A flow can be controlled using an electric field at micro/nanoscale. The physics behind this comes from Electrokinetics. Due to surface chemistry, the majority of solid surfaces have a tendency to acquire a net surface charge that is either positive or negative when in contact with a polar solvent. We exploit this phenomenon to our advantage to either control flow, translocate particles or generate electrical energy. The presence of charged ions in an aqueous solution is indicated by the space charge density. The surface charge density on the wall causes these ions to segregate. At a steady state, the end reservoir gets rich in counterions (ions having opposite charge to that of microchannel walls), giving rise to a potential difference across the microchannel. This potential difference is known as streaming potential. Electrokinetic energy conversion at the micro/nanoscale has become a major area of interest for researchers. Researchers have successfully demonstrated devices that can extract electrical energy from flow through microchannel [6]–[8]. The main issue in this new renewable energy source is its low energy conversion efficiency. Many attempts are being made in a quest to enhance the magnitude of generated electrical potential. The effect of different influencing parameters on energy generation is still ongoing research.

## 2. LITERATURE REVIEW AND OBJECTIVE

The energy conversion challenge in the micro- and nanoscale has been introduced in research labs all over the world as a result of the rapid growth of microelectromechanical systems (MEMS) technology. Yang and Jing[9] studied numerous micro/nanoscale technologies for energy extraction. In this review, conversion of various forms of energy, a number of micro/nanofluid-enabled energy conversion devices are discussed. To generate renewable energy scientists have made efforts in extracting energy from salt concentration gradient. It has been demonstrated that a promising source of clean and renewable energy is the Gibbs free energy released when river and sea water mix. Reverse Electro Dialysis (RED) is one important method for obtaining electrical energy from this natural salt, Yuhui et al.[7] have proposed a novel method to demonstrate this in the context of nanochannels. A number of water-based energy harvesters have been developed in recent years because of their ease of use, sustainability, and environmental friendliness. The devices have so far needed to periodically add water to generate energy continuously, which makes them impractical to use. In order to generate electricity continuously and autonomously, Jaehyeong et al.[10] created an artificial hydrological cycle in a transpiration-driven electrokinetic power generator (TEPG). Researchers are now aware of the uses for energy-producing systems at the micro- and nanoscale and the need to increase their effectiveness. Generation of streaming potential in microscale is an established research model. Numerous attempts have been made[11]–[14] to extend this model in order to maximize the efficiency of these systems and use them for relevant applications. People have become interested in superhydrophobic surfaces because of a variety of natural occurrences, like the movement of water droplets on lotus leaves. Due to the low energy of these surfaces, this effect occurs. Superhydrophobic surfaces feature fascinating physics and practical uses like self-cleaning[15]. Malekidelarestaqi et al.[16] have used these superhydrophobic surfaces and have shown enhanced generation of electrical energy through electrokinetic effect in a pressure driven microchannel. They have used Navier slip length to include hydrophobicity. We have followed the same approach.

The superhydrophobicity of channel walls was quantified by using Navier's slip length.

$$U_{slip} = L_s \left(\frac{\partial U_s}{\partial n}\right)_{wall} \quad (1)$$

A thorough analysis of electrokinetics in relation to non-Newtonian fluids was presented by Zhao and Yang[17]. The topic includes a wide variety of non-Newtonian electrokinetic effects. Shear-thickening liquids tend to reduce electrokinetic effects while shear-thinning liquids tend to enhance electrokinetic phenomena. Additionally, they have offered suggestions for future research directions and addressed a number of non-Newtonian electrokinetics theory-related challenges. In order to highlight the changes in flow dynamics brought on by the interplay between rheology and electrokinetics, several constitutive models that predict the relationship between shear stress and shear strain have been discussed in the literature[18]–[21]. Researchers[22], [23] have also employed a PEL grafted microchannel, which ensures a higher charge density due to the presence of PEL ions and a long slip length, for enhancing power generation. Paul et al.[24] extended this work to highlight the interaction of ion packaging in charged PEL channels, and fluid non-Newtonian constitutive behaviour towards changing the electrokinetics of power-law fluids.

Additionally, most of microfluidic systems operate bio fluids which can't be handled as Newtonian. To describe the flow properties of such fluids, the Navier-Stokes equation should be replaced with the more general Cauchy momentum equation with a relevant constitutive law. Power-law constitutive law is the most basic also widely accepted of the several constitutive laws for non-Newtonian fluids. It has been demonstrated that it is suited for a variety of non-Newtonian fluids, including blood solutions and polymeric solutions. The power law viscosity is given as,

$$\mu_a = m(2\tau)^{(n-1)} \quad (2)$$

Where,

$$\tau = \frac{[\vec{\nabla u} + (\vec{\nabla u})^T]}{2} \quad (3)$$

However, to the best of our knowledge, no attempts were yet made to study the coupled effect of having variable fluid rheology inside a microchannel with superhydrophobic walls for efficiently generating electrical energy sustainably. The main objective of studying this coupled effect is to improve magnitude of generated streaming potential. We have known that solely both the physics can help to improve power generation. Through this study we aim to quantify this improvement when both of the physics are employed. At the same time the study can reveal important insights about effectiveness of one physics over the other in different circumstances.

## 3. PROBLEM STATEMENT

### 3.1 Geometry

The model geometry consists a microchannel with radius 'a' connecting two cylindrical reservoirs with radius that is 5 times 'a'. Figure 1 illustrates a simple 2D geometry with axial symmetry about line AH. Aqueous electrolyte solution is used to fill the reservoirs and channel. The rheology of this fluid is considered to be non-Newtonian which obeys power law. The walls of the microchannel are superhydrophobic with a slip length 'Ls' and they also have a surface charge density. The fluid flows from inlet AB to outlet GH because of a constant pressure difference. Including reservoirs along with the microchannel insures appropriate development at the ends of the microchannel. The bulk conditions are adequately distant from the inlet and the outlet of the microchannel to have almost no influence on the results.

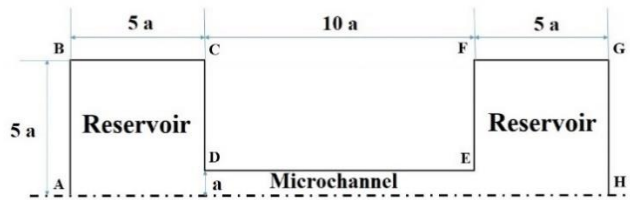

**Figure 1: Schematic of model geome**

### 3.2 Governing Equations

The electrokinetic flow in pressure driven microchannel was modelled with the help of momentum equations for solving velocity field, Poisson's equation for solving electric potential and Nernst-Planck equation for concentration distribution of individual ions across the microchannel.

$$0 = -\nabla p + \nabla [2\mu_a \tau] - \rho_f \nabla \psi \qquad (4)$$

$$\nabla^2 \psi = -\frac{\rho_f}{\varepsilon} \quad \text{where,} \quad \rho_f = \sum e z_i c_i \tag{5}$$

$$0 = \nabla \cdot \left[ c_i \vec{u} - D_i \nabla c_i - \frac{z_i e c_i D_i}{k_B T} \nabla \psi \right] \tag{6}$$

### 3.3 Boundary Conditions

For momentum equation, the boundary AB is inlet with constant pressure $P_0 = 96.15$ Pa and boundary GH is the outlet with 0 Pa static pressure. The no-slip boundary condition was applied for CD and EF. Microchannel walls (boundary DE) is modified with varying wettability which is quantified by the parameter slip length, 'Ls'. Zero electric potential was applied on inlet AB as a gauge boundary condition and a constant surface charge density was applied on wall DE. All other boundaries were kept at zero potential gradient. At last, in order to solve the Nernst-Planck equation, the inlet and outlet walls were assigned bulk concentrations equal to $c_0$, while the remaining walls were given zero flux boundary conditions. Axial symmetry was imparted to the boundary AH.

### 3.4 Solution Methodology

All the listed governing equations (equations 4-6) were solved using commercial software package COMSOL. Solution involved segregated solvers. In the first solution Poisson's equation was solved simultaneously with Nernst-Planck equation. The steady state solution of these equations were used as initial values for the creeping flow module. All the governing equations were solved simultaneously, which gave the concentration variation along the microchannel length. The difference between the potential values at end reservoirs gave us the value of streaming potential.

### 4. RESULTS AND DISCUSSION

In the present study, we have considered

$P_0 = 96.15$ Pa, $k = 1.04 \times 10^7 \text{m}^{-1}$, $e = 1.6021 \times 10^{-19}$ C, $k_B = 1.3806\text{E-}23$ J/K, $a = 5k$, $L = 10a$, $b = 5a$, $T = 298$ K, $D = 10^{-9} \text{m}^2/\text{s}$, $\rho = 1000$ kg/m$^3$, scd = $1.9 \times 10^{-4}$ C/m$^2$, $C_0 = 0.01$ mol/m$^3$.

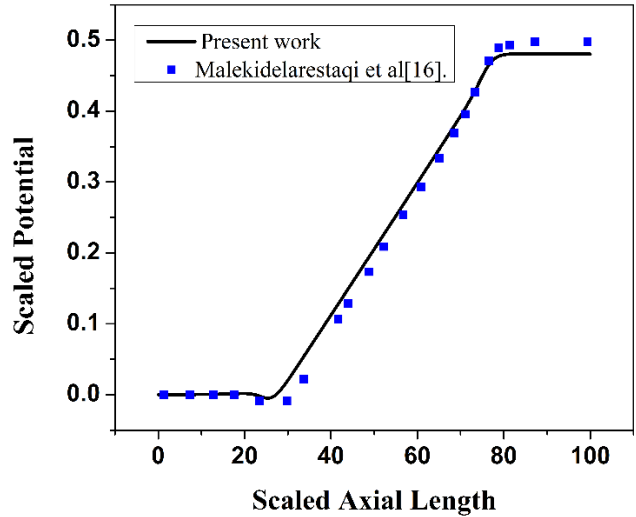

**Figure 2: Model Validation**

To check the accuracy of the present model we have validated our model with the existing results of Malekidelarestaqi et al.[16] considering similar conditions. The present model includes non-Newtonian behaviour of the fluid along with modified channel walls. For validation, we considered n=1 and no slip boundary conditions which is for the case of Newtonian fluid with no modification on channel surface. It is observed from fig. 4 that our results are in good match with the results of Malekidelarestaqi et al.[16]

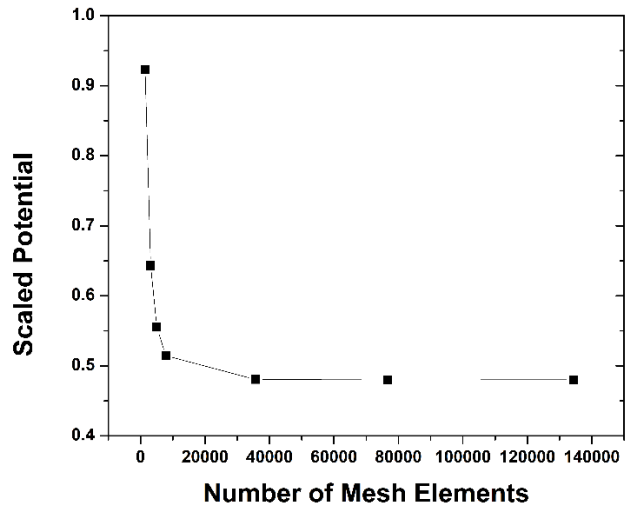

**Fig. 3 Mesh independence study**

All the performed simulations are checked for grid independence. We employed free triangular mesh with varying number of elements. From fig.5 it is observed that the variation in scaled potential is almost negligible beyond 35000 mesh elements and hence, this mesh is considered for further simulations.

We begin with plotting electrical potential with axial length at the centreline of the channel for different flow behaviour indices indicating different fluid rheology. The potential and axial length were non-dimensionalized by multiplying ($e/k_BT$) and k, respectively. The curve corresponding to n = 0.8 describes shear thinning behaviour, n = 1 is for Newtonian fluid while n = 1.2 represents shear thickening nature. It is observed from fig.4 that potential increases along the flow direction due to accumulation of mobile counterions in the downstream reservoir which are dragged with the fluid. It is also evident that with an increase in flow behaviour index the value of potential decreases. This is attributed to the fact that, with decreasing flow behaviour index, the apparent viscosity decreases which results in increase of flow velocity as can be seen in fig 5. This increased velocity drags more number of counterions in the downstream reservoir, thereby enhancing the potential. It is also evident that the magnitude of scaled potential in the downstream reservoir (kz =100) is enhanced 5 times by employing a shear thinning fluid (n = 0.8) as compared to Newtonian fluid (n=1), whereas for shear thickening fluid (n = 1.2) the magnitude reduces by 3.6 times.

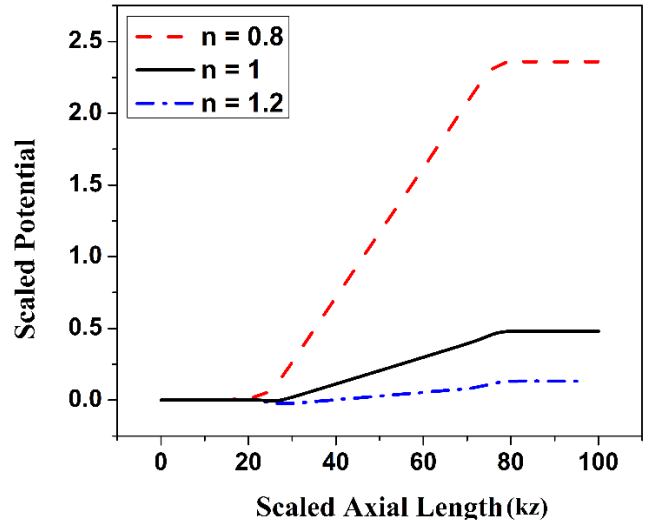

**Figure 4: Axial Variation of Electrical Potential considering no slip, Ls= 0nm**

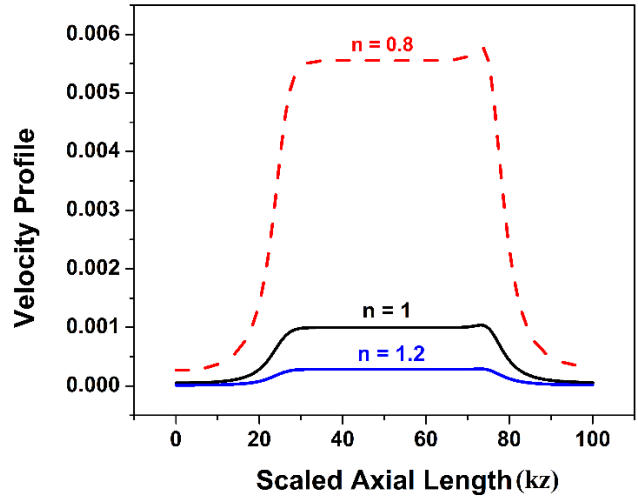

**Figure 5: Axial Variation of Velocity (Ls = 0nm)**

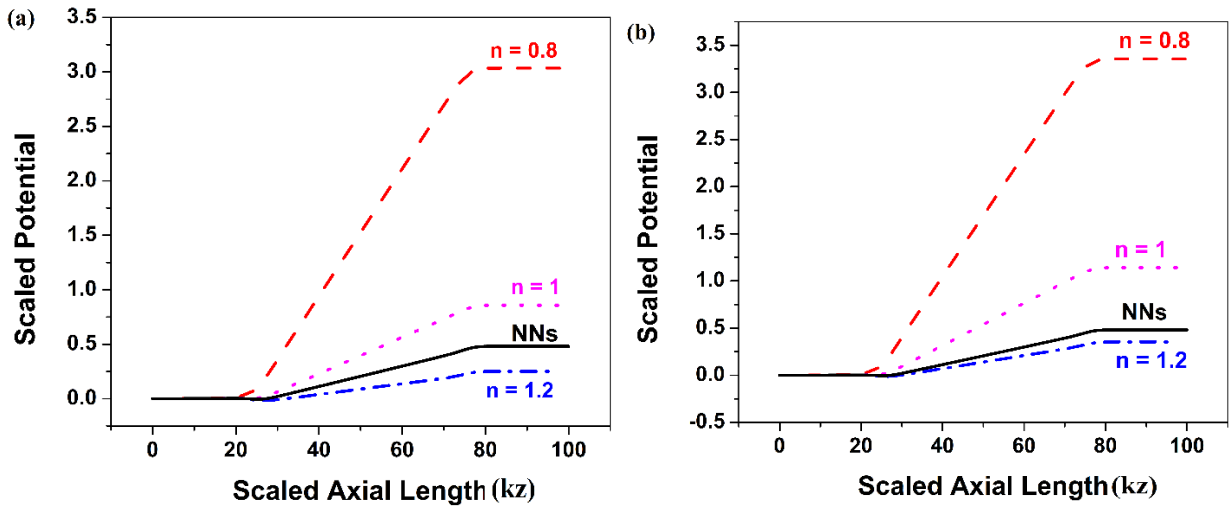

**Figure 6: Axial Variation of Electrical Potential for different flow behaviour index considering (a) slip length, Ls=72 nm, and (b) slip length, Ls = 144nm. The solid line in both figures represents Newtonian fluid without surface modification.**

Figure.6 shows the effect the modified channel surfaces (i.e. slip) on potential variation. The modified surfaces are modelled using Navier slip model (equation 1) with slip length ranging from 0 to 144 nm. Figures 6a and 6b depicts scaled potential variation along the microchannel length considering Ls = 72nm and 144nm, respectively. It is observed that the potential enhances for modified surfaces. i.e. with increase in slip length irrespective of fluid rheology. This is because, due to modified surfaces the mobile counterions very near to the wall are also dragged with the fluid to the downstream reservoir consequently increasing the potential. It can be noted that for the case of Newtonian fluid the potential increases by 77 % for Ls = 72nm (see fig 6a)and by 137 % for Ls = 144 nm(see fig 6b).

Now, to illustrate the combined effect of fluid rheology and surface modification, we compared the potential variation of various non-Newtonian fluid with Newtonian fluid without surface modification abbreviated as NNs (Newtonian,and No-slip) (See fig 6a and 6b). It is observed that for the case of shear thinning fluid (n = 0.8) the potential increases by 530 % for Ls = 72 nm (see fig 6a) and by 600 % for Ls = 144 nm (see fig 6b). as compared to NNs. However, for the shear thickening fluid (n = 1.2) the potential value decreases by 48 % for Ls = 72 nm (see fig 6a) and by 27 % for Ls = 144 nm.(see fig 6b).

Next, we obtained streaming potential as a function of slip length for various fluid behaviour indices. (varying from n = 0.7 to 1.3) shown in figure.7. It is evident that with increasing slip length the streaming potential increases. However, this increment in streaming potential with slip length is insignificant for the case of n = 0.7. This is attributed to higher flow rate at same pumping power due to very low apparent viscosity of fluid (n=0.7). Because of comparatively lower viscosity and higher flow rate the additional effect of slip on flow rate is insignificant. It is observed that, the streaming potential decreases with increase in n. Also, the variation of streaming potential with flow behaviour index is very less for shear thickening fluid (i.e. for n > 1) as compared to the shear thinning fluid (i.e. for n < 1). This is attributed to the flow velocity. With decrease in n (from 1 to 0.8), the flow velocity increases by 5.5 times whereas, with increase in n (from 1 to 1.2), the flow velocity reduces by 3.5 times (see figure.5).

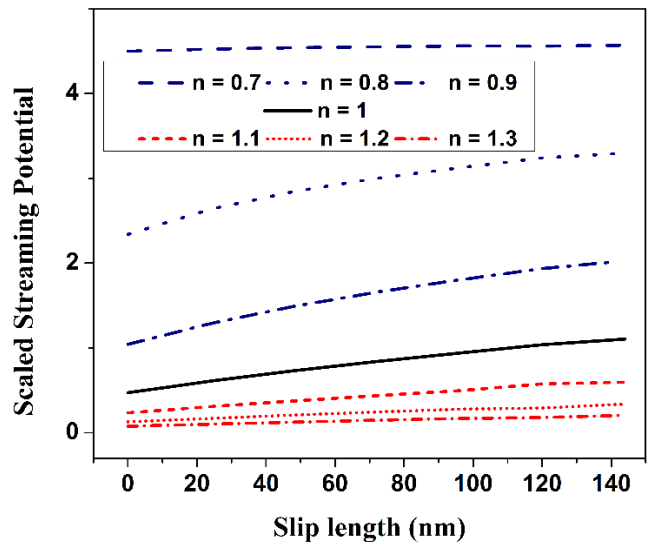

**Figure 7: Scaled Streaming potential vs Slip length for varying flow behaviour index, n.**

In figure 8a,8b and 8c, radial variation of streaming current is illustrated. The scaled streaming current is calculated as $\frac{U}{kD}(\frac{c_1-c_2}{c_0})$. It is observed that for no slip the streaming current attains a maximum value at kr=4 (Debye length from the wall) for all fluids. However, with increasing slip the maximum streaming current will shifts towards walls. Also, the value of streaming current enhances with an increase in slip. This is because as wall slip increases more ions just next to the microchannel wall are dragged along the fluid flow consequently increases the streaming current near the walls. It is also evident that for the case of shear thinning fluids (n<1) the value of streaming current is higher whereas for shear thickening fluids (n>1) the value of streaming current is lower than Newtonian fluids. This is attributed to the higher flow velocity for the former case and lower flow velocity for the latter case than the Newtonian fluid (see figure.5).

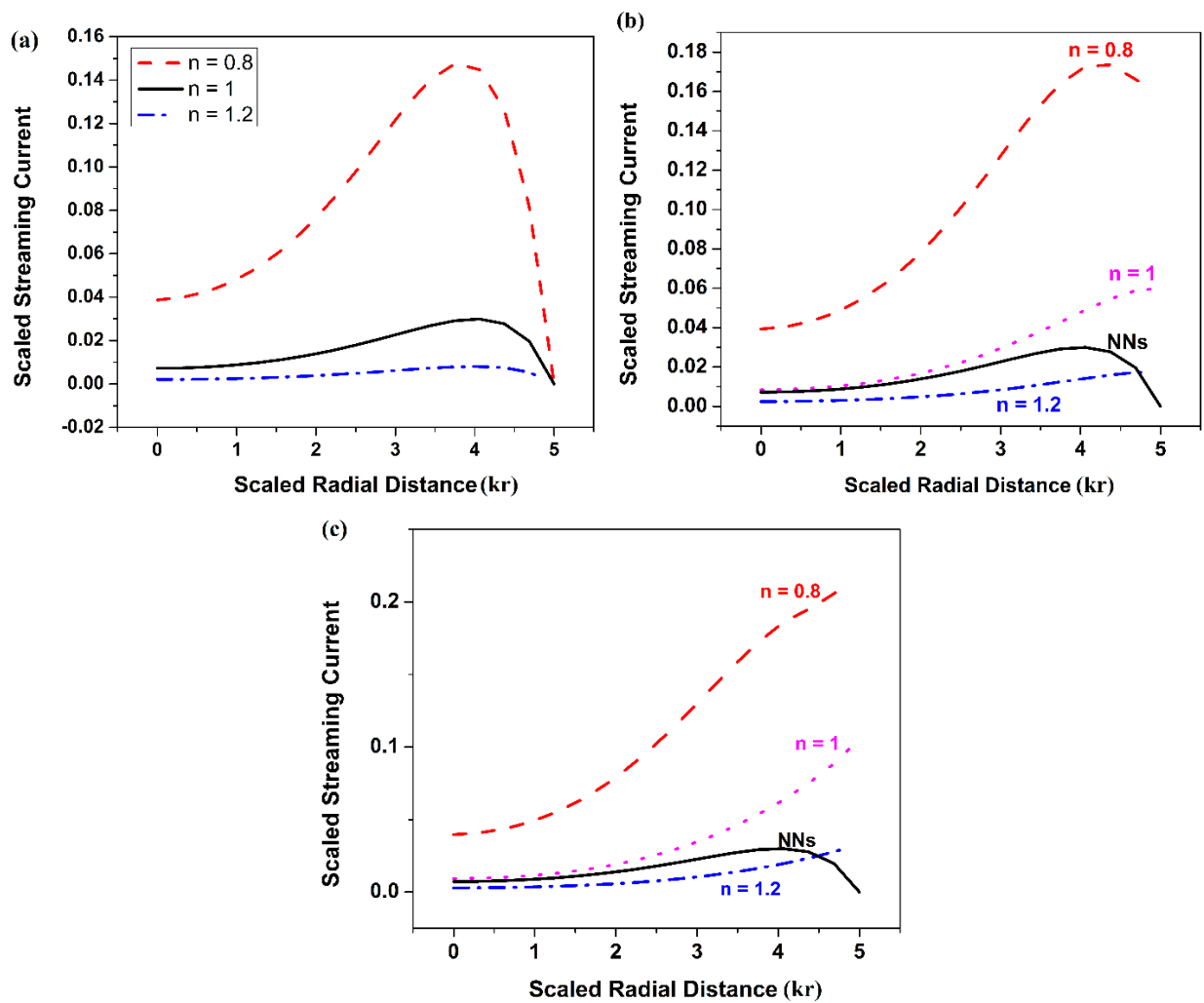

**Figure 8: Radial Variation of Scaled Streaming Current for different flow behaviour index considering (a) no slip (b) slip length, Ls=72 nm, and (c) slip length, Ls = 144nm. The solid line in both figure represent Newtonian fluid without surface modification.**

## 5. CONCLUSIONS

A numerical study of electokinetic energy generation through a modified channel with non-newtonian fluid has been performed. The modified channel with varying wettability is quantified as varying slip lengths ranging from 0 to 144 nm. The non-newtonian fluid is modelled using power-law, where flow behaviour index, n is varied from 0.7 to 1.3. The newtonian fluid is represented by n=1, whereas n>1 and n<1 represents shear thickening and shear thinning fluids, respectively. We demonstrated that through channel modification which leads to slip at fluid-solid interface, the streaming potential and streaming current can be enhanced irrespective of fluid rheology. Further, we studied the combined effect of fluid rheology and modified channel. It can be concluded that the streaming potential and streaming current get enhanced for shear thinning fluid and reduced for shear thickening fluid as compared to Newtonian fluid. We demonstrated that the maximum enhancement in streaming potential is 850 % for n=0.7 considering Ls=144 nm whereas maximum reduction in streaming potential is 60 % for n=1.3 considering the same slip length. For lower values of slip length, the streaming current attains a maximum value, but as wall slip increases, the streaming current curve gets more flat near the walls. The maximum enhancement in streaming current by 650 % for n=0.8 considering Ls=144 nm. With these insights, section of fluid and substrate can be done based on a mapping of power law index and slip length to optimize the magnitude of streaming potential. For lower values of power law index having superhydrophobic surface does not aid significantly at the same time the jump observed in streaming potential for shear thinning fluids is comparatively much higher than the dip visualised for shear thickening fluids, these are intriguing insights reported in this study. These findings will guide towards the development of better nanofluidic devices with non-Newtonian fluids and wall hydrophobicity

## NOMENCLATURE

| | | |
|---|---|---|
| $L$ | Length of microchannel | [m] |
| $b$ | Height of reservoir | [m] |
| $a$ | Radius of microchannel | [m] |
| $Ls$ | Slip length | [m] |
| $\mu_a$ | Power law viscosity | [Pa.s] |
| $\Psi$ | Electric Potential | [V] |
| $\rho_f$ | Space charge density | [C/m$^3$] |
| $e$ | Protonic charge | [C] |
| $z$ | Valence of ions | -- |
| $n$ | Ion concentration | [mol/m$^3$] |

| $p$ | Pressure | [Pa] |
| $u$ | Velocity | [m/s] |
| $\epsilon$ | Permittivity | [F/m] |
| $D$ | Diffusion constant | [m²/s] |
| $k_B$ | Boltzmann constant | [J/K] |
| $T$ | Temperature | [K] |
| $\tau$ | Shear stress | [Pa] |
| $m$ | Flow consistency index | [Pa.s] |
| $n$ | Flow behaviour index | -- |
| $k$ | Inverse of De-Bye length | [1/m] |